\documentclass[aps,twocolumn,prl,showpacs]{revtex4}
\usepackage{graphicx,amssymb,amsmath}
\usepackage{graphicx}
\usepackage{dcolumn}
\usepackage{bm}

\newcommand{\nt}{$\nu_T=1$}
\newcommand{\dl}{$d/\ell$}
\newcommand{\dlc}{$(d/\ell)_c$}
\newcommand{\ez}{$E_Z$}
\newcommand{\rxxd}{$R_{xx,D}$}
\newcommand{\rxyd}{$R_{xy,D}$}
\newcommand{\ddl}{$\Delta (d/\ell)$}

\begin{document}

\title{Quantum Hall Exciton Condensation at Full Spin Polarization}

\author{A.D.K. Finck$^1$, J.P. Eisenstein$^1$, L.N. Pfeiffer$^2$, and K.W. West$^2$}

\affiliation{$^1$Condensed Matter Physics, California Institute of Technology, Pasadena, CA 91125
\\
$^2$Department of Electrical Engineering, Princeton University, Princeton, NJ 08544}

\date{\today}

\begin{abstract}Using Coulomb drag as a probe, we explore the excitonic phase transition in quantum Hall bilayers at \nt\ as a function of Zeeman energy, \ez. The critical layer separation \dlc\ for exciton condensation initially increases rapidly with \ez, but then reaches a maximum and begins a gentle decline. At high \ez, where both the excitonic phase at small \dl\ and the compressible phase at large \dl\ are fully spin polarized, we find that the width of the transition, as a function of \dl, is much larger than at small \ez\ and persists in the limit of zero temperature. We discuss these results in the context of two models in which the system contains a mixture of the two fluids.
\end{abstract}

\maketitle
Following the development of the Bardeen-Cooper-Schrieffer theory of superconductivity, physicists \cite{blatt,moskalenko,keldysh,lozovik,shevchenko} speculated that excitons in semiconductors, conduction band electrons bound to valence band holes, could undergo a similar pairing transition to a collective state with macroscopic quantum phase coherence. Strong evidence of exciton condensation was ultimately found in a surprising place: double layer two-dimensional electron systems at high perpendicular magnetic field $B_\perp$ \cite{spielman1,kellogg1,kellogg2,tutuc1,wiersma}.  In this, the quantum Hall effect regime, excitons consisting of electrons in the lowest Landau level (LLL) of one layer bound to holes in the LLL of the other layer, condense into a coherent collective state whenever the temperature and layer separation are small enough, and the total density $n_T$ of electrons in the double layer system equals the degeneracy $eB_\perp/h$ of one spin-resolved Landau level \cite{fertig,macd1,yoshioka,wen}. This collective electronic state exhibits several dramatic electrical transport properties including Josephson-like interlayer tunneling \cite{spielman1}, quantized Hall drag \cite{kellogg1} and vanishing Hall resistance \cite{kellogg2,tutuc1,wiersma} when currents are driven in opposition in the two layers. 

Exciton condensation in bilayer quantum Hall systems at total Landau level occupancy $\nu_T \equiv n_T/(eB_\perp/h) = 1$ reflects a spontaneously broken U(1) symmetry in which electrons are no longer confined to one layer or the other but instead reside in coherent linear combinations of the two.  This interlayer phase coherence develops only when the effective interlayer separation \dl\ (with $d$ the center-to-center quantum well separation and $\ell = (\hbar/eB_\perp)^{1/2}$ the magnetic length) is less than a critical value, \dlc. At large \dl\ the bilayer system behaves qualitatively like two independent two-dimensional electron systems (2DESs).  The nature of the quantum phase transition separating these two very different bilayer states remains poorly understood.  In particular, while essentially all theoretical work concerning the transition makes the simplifying assumption that both phases are fully spin polarized \cite{fertig,macd1,yoshioka,wen,yang1,moon,schliemann,stern,simon,moller}, recent experiments \cite{spielman2,kumada} convincingly demonstrate that this is not the case in typical samples.  Instead, these experiments prove that the two phases have different spin polarizations, with the polarization of the excitonic phase at small \dl\ exceeding that of the incoherent phase at large \dl. (The data further suggest that the polarization of the excitonic phase is in fact complete, at least in the zero temperature limit.) While a cleanly observed jump in spin polarization at a critical \dl\ would strongly suggest that the phase transition is first order, in the actual samples the transition appears to be continuous.  Whether this is the result of a fundamentally first-order transition smeared by disorder \cite{stern,rick} or is an intrinsic attribute \cite{simon,moller} of strongly correlated bilayers at \nt\ remains unknown. 

Using tilted magnetic fields to increase the electronic Zeeman energy $E_Z$, Giudici {\it et al} \cite{Giudici} have recently been able increase the spin polarization of the incoherent phase and to approach the situation where both phases are fully polarized. The expected \cite{spielman2,kumada} increase of \dlc\ with $E_Z$ was clearly detected \cite{Giudici}, but no qualitative change in the character of the excitonic phase transition was reported.  Here we report the observation of just such a qualitative change in the character of the transition as both phases become fully spin polarized.  Our data show that the transition is dramatically broadened when both phases are spin polarized and that this broadening persists in the limit of zero temperature.  

Our results are based on a study of Coulomb drag \cite{gramila}, the force one current-carrying 2DES exerts on a second, electrically isolated 2DES, in a GaAs/AlGaAs double quantum well (DQW). The DQW consists of two 18 nm GaAs quantum wells separated by a 10 nm Al$_{0.9}$Ga$_{0.1}$As barrier (and hence $d=28$ nm).  As grown, remote Si donors populate each quantum well with a 2DES with a density and low temperature mobility of $n \approx 5.5 \times 10^{10}$ cm$^{-2}$ and $\mu \approx 1 \times 10^6$ cm$^2$/Vs, respectively.  The sample is patterned into a 250 $\mu$m square with narrow arms extending outward from each side to diffused NiAuGe ohmic contacts.  A selective depletion scheme allows these contacts to connect separately to one or the other 2DES in the central square region.  By tilting the sample relative to an external magnetic field $B_T$, the Zeeman energy $E_Z=g \mu_B B_T$ can be adjusted at the fixed perpendicular field $B_\perp$ needed to establish $\nu_T = 1$. The crucial ratio \dl\ ($=d\sqrt{2\pi n_T}$ at \nt) is tuned {\it in situ} via gate electrodes which control the individual layer densities $n_{1,2}$; we focus here on the balanced, $n_1=n_2=n_T/2$ case. Drag measurements are performed by driving a small ac current $I$ (typically 0.5 nA at 13 Hz) through two contacts on one layer while measuring the induced voltage $V_D$ between two contacts on the other layer; the drag resistance is defined as the ratio $V_D/I$.  As expected, the measured drag resistance is unchanged when the role of the two layers is interchanged. 

\begin{figure}
\includegraphics[width=3.1in, bb=155 204 420 425]{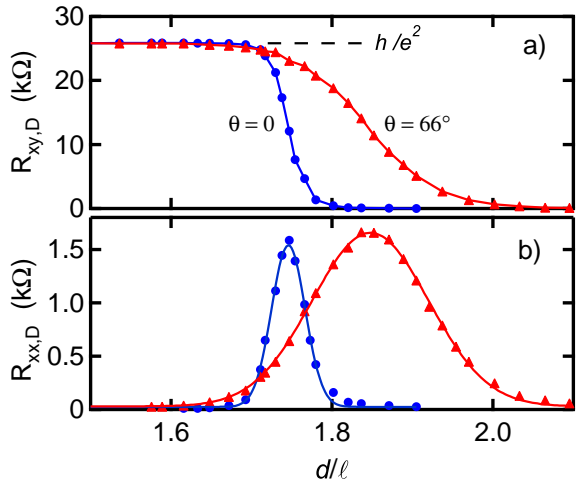}
\caption{\label{}(color online) Hall and longitudinal Coulomb drag at \nt\ vs. \dl\ at $T = 30$ mK. Dots, $\theta = 0$; triangles, $\theta = 66^\circ$. Solid curves in b) are gaussian fits.}
\end{figure}

Figure 1 shows the measured Hall and longitudinal drag resistances, \rxyd\ and \rxxd, respectively, at \nt\ {\it vs.} \dl\ at $T = 30$ mK.  Two sets of data are shown, one with the sample plane perpendicular to the magnetic field ($\theta = 0$) and one with the sample normal rotated by $\theta = 66^\circ$ relative to the field.  As observed previously \cite{kellogg1} at $\theta = 0$, \rxyd\ rises from zero at large effective layer separation \dl\ to the quantized value \rxyd\ = $h/e^2$ at small \dl.  The passage between these two values is relatively sharp in \dl\ and signals the transition between the compressible bilayer \nt\ phase at high \dl\ and the excitonic phase at small \dl.  Quantization of Hall drag is expected \cite{moon,yang2} in the excitonic phase and has been elegantly explained via an analogy to a superconducting Giaever flux transformer \cite{girvin}. Coincident with the step in \rxyd, the longitudinal drag resistance \rxxd\ exhibits a strong peak \cite{sign}.  While various models \cite{stern,simon} have been advanced to explain the origin of this peak, for the moment it suffices to employ it to define the location \dlc\ \cite{midpoint} and width \ddl\ of the transition between the excitonic and compressible phases at \nt.

Figure 1 also demonstrates that the same basic transition is observed when the sample is tilted by $\theta = 66^\circ$. This is not unexpected since it is well known that the quantized Hall effect at \nt\ in conventional transport is robust against the application of in-plane magnetic fields \cite{murphy}. While this proves the existence of an energy gap, the quantization of Hall drag further demonstrates that non-perturbative interlayer correlations are central to the structure of the phase \cite{yang2}. Beyond the similarities of the data sets at $\theta = 0$ and $66^\circ$, there are two important differences: tilting the sample shifts the transition region from $d/\ell \approx 1.75$ to $d/\ell \approx 1.85$, and substantially broadens it. This broadening is equally evident in the Hall and longitudinal components of the drag.

\begin{figure}
\includegraphics[width=3.1in, bb=0 0 265 225]{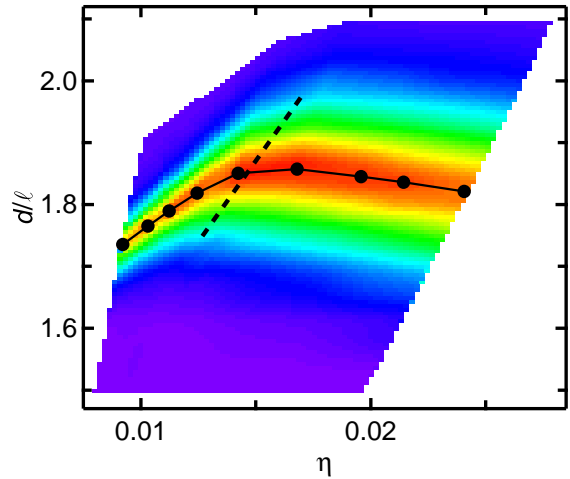}
\caption{\label{}(color online) Longitudinal drag \rxxd\ {\it vs.} \dl\ and $\eta=E_Z/(e^2/\epsilon\ell)$ at \nt\ and $T = 50$ mK. Color scale: Purple, $R_{xx,D}=0$; Red, $R_{xx,D}=2$ k$\Omega$. Solid dots: Phase boundary, \dlc\ {\it vs.} $\eta$. Dashed line: Approximate location of ``knee'' in drag contours.  Left and right boundaries of colored region: $\theta = 0$ and $\theta = 66^\circ$ trajectories, respectively.}
\end{figure}
Figure 2 displays a color map of the longitudinal drag \rxxd\ at \nt\ and $T = 50$ mK, versus \dl\ and $\eta = E_Z/(e^2/\epsilon \ell)$, the Zeeman energy normalized by the mean intra-layer Coulomb energy $e^2/\epsilon \ell$.  The plot is constructed from 9 sets of \rxxd\ $vs.$ \dl\ measurements at tilt angles ranging from $\theta = 0$ to $\theta = 66^\circ$. (Note that for fixed $\theta$, varying \dl, by changing $n_T$, at \nt\ traces out a straight-line trajectory $d/\ell \propto \eta$, with slope proportional to $\cos{\theta}$. Trajectories for $\theta = 0$ and 66$^\circ$ form the left and right boundaries, respectively, of the colored region in Fig. 2.)  There are several important features in Fig. 2.  First, at each $\theta$ \rxxd\ exhibits a well-defined peak $vs.$ $d/l$.  As the coloring suggests, the height of this peak is roughly independent of $\eta$, varying by no more than $\pm 5$\% from its average value. We use the center of the peak to define the critical effective layer separation \dlc\ between the excitonic and incoherent phases; these determinations are indicated by the solid dots in the figure \cite{midpoint}. As $\theta$ is first increased from zero, thus increasing $\eta$, \dlc\ rises steadily.  This previously observed effect \cite{spielman2,kumada,Giudici} is due to the difference in spin polarization $\Delta \xi = \xi_{ex} - \xi_i$ of the excitonic and incoherent phases at \nt. Since $\xi_{ex}>\xi_i$ at small $\eta$ \cite{spielman2,kumada}, tilting the sample and thereby increasing \ez\ enhances the stability the excitonic phase, and thus increases \dlc. As $\eta$ is further increased, the energetic advantage of the excitonic phase declines owing to the increasing spin polarization of the incoherent phase.  Once full spin polarization of the incoherent phase is reached, \dlc\ should cease changing with $\eta$.  This is the regime approached, but not entered, by Giudici {\it et al.} \cite{Giudici}. Here we reach higher values of $\eta$ and thereby probe deeply into the fully polarized regime.

The data in Fig. 2 show that this simplest scenario is not quite observed.  After a steady rise, \dlc\ reaches a maximum and then begins a gentle decline with $\eta$.  We believe that this decline is in fact an {\it orbital} effect of the large in-plane magnetic field and that the two phases are indeed fully spin polarized in this regime. In addition to enhancing \ez, tilted magnetic fields lead to level mixing effects due to the finite thickness of the 2D layers. In the present symmetrically doped DQW, these mixing effects lead to ``squashing'' of the subband wavefunction and an increase in the mean separation between electrons in the two layers.  Both effects reduce the relative stability of the excitonic phase, and we believe they are responsible for the slow decline of \dlc\ at large $\eta$.

A second surprising aspect of the data in Fig. 2 concerns the location of the ``knee'' in the various drag contours.  The dashed line in the figure represents the approximate location of this knee.  If the knee indicates the onset of full spin polarization of the incoherent phase, then the finite slope of the dashed line implies that the normalized Zeeman energy $\eta$ required for full polarization depends on the effective layer separation \dl, at least within the transition region.  If within this transition region the bilayer 2DES is in some kind of mixed phase (either heterogeneously \cite{stern} or homogeneously \cite{simon,moller}), with attributes of both the excitonic and incoherent fluids, then the finite positive slope of the dashed line suggests that the incoherent fluid polarizes more readily when spin polarized excitonic fluid is also present.

Finally, Fig. 2 clearly displays the broadening of the transition region noted above. The width of the transition appears to increase steadily while \dlc\ is rising with $\eta$. At higher $\eta$, beyond the knees in the drag contours, the width of the transition appears to saturate.  This is made clearer in Fig. 3(a) where the width \ddl, defined as the half width at half maximum of the \rxxd\ peak, at $T = 50$ mK is plotted {\it vs.} $\eta$.  The total change in the width is quite substantial: about a factor of 3. 

\begin{figure}
\includegraphics[width=3.1in, bb=151 213 435 459]{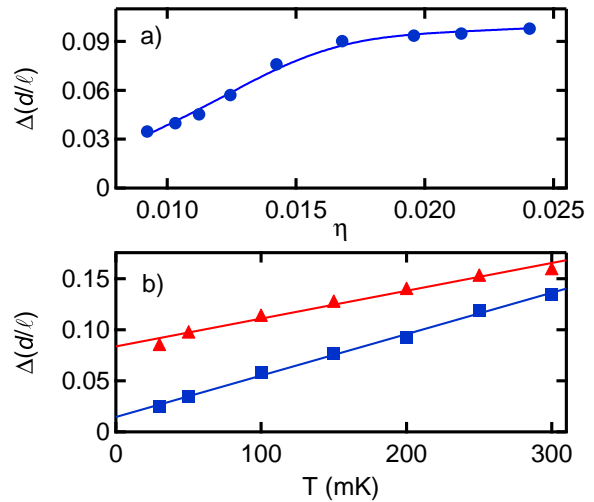}
\caption{\label{}(color online) a) Width, \ddl, of the longitudinal drag peak at \nt\ and $T=50$ mK versus normalized Zeeman energy $\eta$ at the peak center.  b) Temperature dependence of \ddl\ at $\theta = 0$ (blue squares) and $\theta = 66^\circ$ (red triangles).}
\end{figure}
The large increase in \ddl\ is not simply due to an enhancement of thermal fluctuations.  Figure 3(b) shows the temperature dependence of \ddl\ at $\theta = 0$ and $\theta = 66^\circ$.  As observed previously \cite{kellogg3}, at $\theta = 0$ \ddl\ increases linearly with $T$ but is very small in the $T \rightarrow 0$ limit.  In contrast, at $\theta = 66^\circ$, in the spin polarized regime, \ddl\ clearly does not vanish in the zero temperature limit.  This implies the presence of a non-thermal broadening mechanism.

Stern and Halperin (SH) \cite{stern} attribute the peak in \rxxd\ to phase separation near the critical \dl. The source of this phase separation is assumed to be static fluctuations in the 2DES density arising from spatial inhomogeneities in the Si donor populations in the sample.  For $d/\ell \gtrsim (d/\ell)_c$ the bilayer 2DES is presumed to consist largely of the incoherent phase with small inclusions of the excitonic phase. For $d/\ell \lesssim (d/\ell)_c$ the situation is reversed.  The large \rxxd\ near the critical point results from the very different transport properties of the two phases: while the incoherent phase consists of two essentially independent compressible composite fermion (CF) metals, the excitonic phase is an incompressible quantum Hall conductor for parallel currents in the two layers but a superfluid for antiparallel currents \cite{moon,kellogg2,tutuc1,wiersma}.  Among the predictions of this model is a universal value, $\rho_{xx,D}=h/2e^2$, for the peak drag resistivity.

Although the SH model was constructed assuming full spin polarization of both phases, it should certainly be possible to modify it to include partial polarization of the incoherent phase.  It is unclear to us whether such a modification could account for the much larger transition width in the polarized regime and the apparent dependence of the Zeeman energy $\eta$ required to produce full polarization (the knee) on the layer separation \dl.  One interesting possibility is that since the density of states of the CFs presumably drops by a factor of two upon full spin polarization, their ability to respond to the disorder potential would be changed.  This could result in a different transition width.  

It may also be possible to understand our results without assuming that the phase transition is first order.  Simon, Rezayi, and Milovanovic (SRM) have proposed that as \dl\ is reduced, the bilayer \nt\ system evolves smoothly from an incoherent phase consisting of two independent CF metals to a coherent phase comprised of Bose condensed composite bosons (CB).  At intermediate \dl\ CFs and CBs coexist, and SRM present mixed CF-CB wavefunctions to describe the many-body ground state in this regime. This homogeneously mixed CF-CB model leads to the same drag transport signatures that SH find in their phase separation model. 

A simple mean-field model of the energetics of the SRM scenario may be constructed as follows. We assume the condensed CBs are fully spin polarized, but allow for partial polarization of the CFs.  Let $f$ denote the fraction of electrons in the CF phase.  To allow for partial polarization of the CFs, we write $f=f_{\uparrow}+f_{\downarrow}$.  Treating the CFs and CBs as independent, the total energy per electron in the \nt\ bilayer system is assumed to be $E=\frac{1}{2}E_{F0}(f^2_{\uparrow}+f^2_{\downarrow}) - \frac{1}{2}E_Z(f_{\uparrow}-f_{\downarrow}) + (1-f)(C-\frac{1}{2}E_Z)$.  Here $E_{F0}$ is the Fermi energy of the system assuming it consisted only of spin polarized CFs \cite{mstar}, and $C$ is the energy of each condensed CB. Minimizing $E$, we find $f_{\uparrow}=C/E_{F0}$ and $f_{\downarrow}=(C-E_Z)/E_{F0}$.  Thus, if $E_Z<C$, CFs of both spins are present and $f=(2C-E_Z)/E_{F0}$. Conversely, if $E_Z>C$, the CFs are fully polarized ($f_{\downarrow}=0$) and $f=C/E_{F0}$. Clearly, a mixed phase ($0<f<1$) will exist over a range of $C$ in both the partially and fully polarized CF regimes. In the partially polarized regime, contours of fixed $f$ satisfy $C=(f E_{F0} + E_Z)/2$ and thus rise linearly with $E_Z$.  However, when the Zeeman energy reaches $E_Z = C$, the CF phase becomes fully spin polarized ($i.e.$ $f_{\downarrow}=0$) and the contour becomes independent of Zeeman energy.  The ``knee'' in the contour occurs when $E_Z=fE_{F0}$ and is thus proportional to the CF fraction, $f$. If we use the range $1/4 < f < 3/4$ to define the transition width $\Delta C$, we find that the width $\Delta C = E_{F0}/ 2$ in the fully polarized regime is twice as large as $\Delta C = E_{F0}/4$ in the partially polarized regime. Making the plausible assumption that the condensed CB energy $C$ increases linearly with $d/\ell$ near the critical point, these conclusions agree qualitatively with all of the main features of the drag contours in Fig. 2, although we stress that a theory directly relating $f$ to the measured drag resistances is lacking.

In summary, we have observed how quantum Hall exciton condensation in bilayer systems depends on Zeeman energy and have probed deeply into the fully polarized regime.  Several qualitative effects have been observed, most notably a dramatic increase in the width of the transition when all spins become polarized.  We have discussed our results in terms of two models of the transition, both of which assume mixing of excitonic and composite fermion phases.  It remains to be determined whether this mixing occurs via first order phase separation or reflects a homogeneous cross-over.  

We thank Allan MacDonald, Gil Refael, Steve Simon, Ady Stern, and Yue Zou for helpful discussions.  This work was supported via NSF grant DMR-0552270.


\begin{references}
\bibitem{sign} Since the longitudinal drag voltage is opposite in sign to the longitudinal voltage in the current-carrying layer, we define $R_{xx,D}=-V_{xx,D}/I$.
\bibitem{mstar} We are ignoring any dependence of the CF effective mass on $f$.  But even if we assume $E_F \propto f^{1/2}$, as might be argued, none of our qualitative conclusions are altered.
\end{references}
\end{document}